\documentclass[article]{IEEEtran}
\IEEEoverridecommandlockouts

\usepackage{amsmath,amssymb,capt-of}

\usepackage[utf8x]{inputenc}
\usepackage[super]{nth}

\usepackage{cite}

\usepackage[acronym,toc]{glossaries}

\newcommand{\ie}{i.e.,~}
\newcommand{\eg}{e.g.,~}

\usepackage[inline]{enumitem}

\usepackage[table]{xcolor}

\usepackage{array}

\usepackage{caption}
\usepackage{subcaption}
\usepackage{float}

\usepackage{lastpage,fancyhdr,graphicx}

\usepackage[linesnumbered,lined,boxed,commentsnumbered]{algorithm2e}

\newlength\MAX  

\usepackage{multirow}
\usepackage{tabularx}

\usepackage{etoolbox}
\makeatletter
\patchcmd{\@verbatim}
  {\verbatim@font}
  {\verbatim@font\small}
  {}{}
\makeatother

\usepackage{nameref,hyperref}

\begin{document}

\title{Spotting political social bots in Twitter:\\A use case of the 2019 Spanish general election}

\author{
    \IEEEauthorblockN{Javier Pastor-Galindo\IEEEauthorrefmark{1}, Mattia Zago\IEEEauthorrefmark{1}, Pantaleone Nespoli\IEEEauthorrefmark{1}, Sergio L\'opez Bernal\IEEEauthorrefmark{1}, Alberto Huertas Celdr\'an\IEEEauthorrefmark{2}\IEEEauthorrefmark{3}, Manuel Gil P\'erez\IEEEauthorrefmark{1}, Jos\'e A. Ruip\'erez-Valiente\IEEEauthorrefmark{1}, Gregorio Mart\'inez P\'erez\IEEEauthorrefmark{1}, F\'elix G\'omez M\'armol\IEEEauthorrefmark{1}}\\
    \IEEEauthorblockA{\IEEEauthorrefmark{1}\textit{Department of Information and Communications Engineering, University of Murcia}, 30100 Murcia, Spain
    \\\{javierpg, mattia.zago, pantaleone.nespoli, slopez, mgilperez, jruiperez, gregorio, felixgm\}@um.es}\\
    \IEEEauthorblockA{\IEEEauthorrefmark{3}\textit{Telecommunication Software \& Systems Group, Waterford Institute of Technology}, X91 Waterford, Ireland}\\
    \IEEEauthorblockA{ \IEEEauthorrefmark{2}\textit{Department of Informatics, University of Zurich}, 8050 Zürich, Switzerland
    \\huertas@ifi.uzh.ch}

}

\maketitle

\begin{abstract}
While social media has been proved as an exceptionally useful tool to interact with other people and massively and quickly spread helpful information, its great potential has been ill-intentionally leveraged as well to distort political elections and manipulate constituents. In the paper at hand, we analyzed the presence and behavior of social bots on Twitter in the context of the November 2019 Spanish general election. Throughout our study, we classified involved users as social bots or humans, and examined their interactions from a quantitative (\ie amount of traffic generated and existing relations) and qualitative (\ie user's political affinity and sentiment towards the most important parties) perspectives. Results demonstrated that a non-negligible amount of those bots actively participated in the election, supporting each of the five principal political parties.
\end{abstract}


\section{Introduction}\label{sec:introduction}

Social media have become one of the main channels to spread information worldwide at scale and their popularity renders them as one of the most impactful means of influencing public opinion~\cite{Zhuravskaya2019,GomezMarmol-IC14}.
As stated by the 2019 Global Inventory of Organised Social Media Manipulation report~\cite{cybertroop-report2019} elaborated by the University of Oxford:

\begin{quote}
    \emph{``Social media, which was once heralded as a force for freedom and democracy, has come under increasing scrutiny for its role in amplifying disinformation, inciting violence, and lowering levels of trust in media and democratic institutions.''}
\end{quote}

To this extent, one of the most powerful strategies to maximize the dissemination of a message that aims to deceive social media users consists on using social bots as amplifiers~\cite{Kusen2020}.
Those software-controlled social accounts are able to effectively mimic the normal behavior of human users while sneakily operating at a much higher rate and remaining obscure~\cite{Ferrara2019a}.
In particular, recent studies have disclosed how these coordinated armies are working to poison democratic elections in an orchestrated manner~\cite{Luceri2019a}.

While social media enables the fast propagation of fake news or any other misleading information over the web, the so-called \emph{political social bots} take care of amplifying their popularity to catch the eye of virtual communities and to create manually crafted viral trends~\cite{Yang2019}. 

They share a common characteristic: the abuse of automation tools to generate huge amounts of social media activity in order to support, or oppositely attack, political figures following their agenda with personal interests. Alarmingly, bots are progressively becoming more sophisticated thanks also to the advances in Artificial Intelligence~\cite{Ferrara2019}. Slowly, year by year, bots can build more realistic social media behaviors and produce credible content with human-like temporal patterns~\cite{Stella2019}, while creating a coordinated network to spread the forged information further~\cite{Besel2018}.

\begin{figure}[h]
    \centering
    \includegraphics[width=\columnwidth]{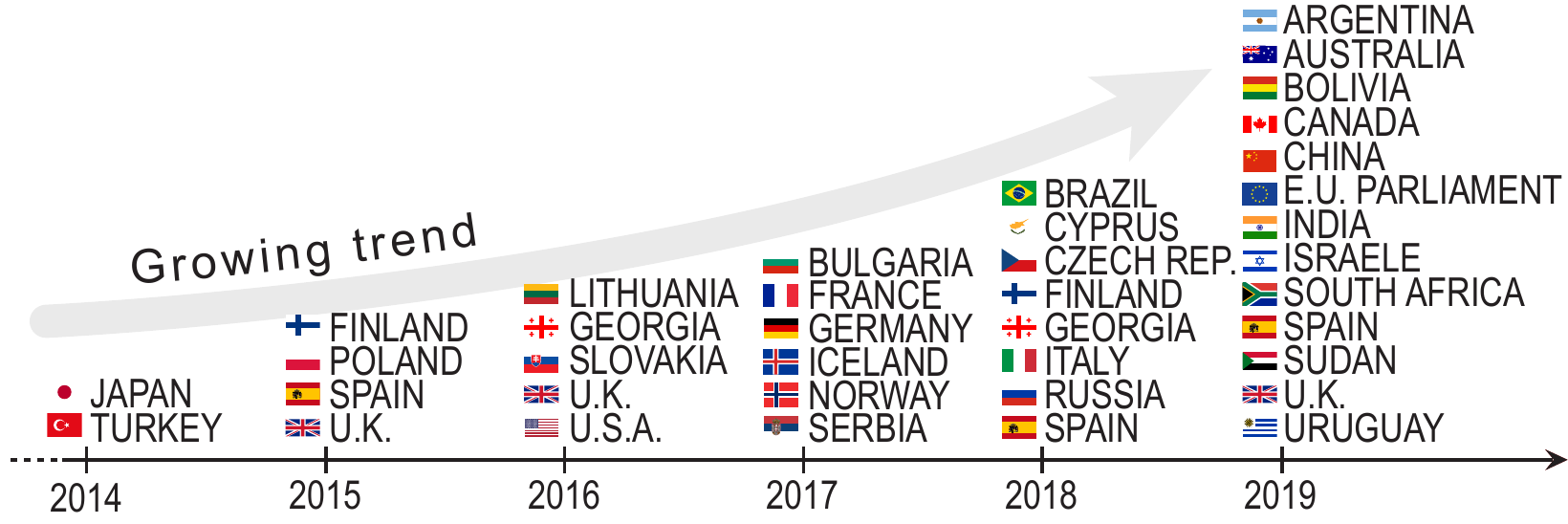}
    \caption{Recent political and administrative elections with social bots participation.}
    \label{fig:5yTimeline}
\end{figure}

In this regard, social bots represent a growing phenomenon (see~\figureautorefname~\ref{fig:5yTimeline}) which aims at jeopardizing modern democracies by distorting reality and manipulating constituents~\cite{Badawy2019}. These malicious operations are often referred to as astroturfing or Twitter bombs, which fake the appearance of organic grassroot participation while being secretly orchestrated and funded~\cite{Keller2019}. In such a scenario, it is clear that this threat is more present and real than ever, potentially affecting millions of users that are absolutely unaware of these malicious activities which may undermine worldwide democracies~\cite{Aral2019}.

Given the above-mentioned non-negligible threats and dangers to our democracies and modern societies, there is an imperative and urgent need to develop innovative solutions to safeguard defenseless citizens from ill-intentioned manipulation.
In this sense, Open Source Intelligence (OSINT)~\cite{Pastor-Galindo2019} becomes a promising paradigm with which to perform an in-depth analysis of publicly accessible sources, such as social networks, and tackle social media manipulation.

One of the major challenges within this larger issue has been to effectively model the behavior of these social bots. This information can greatly help to improve the detection methods and to better quantify the impact that these bots can have on social events and real world decisions. To contribute to address this challenge, in this work we perform an in-depth case study of the activity and behavior of social bots on Twitter in the context of the November 2019 Spanish general election. More specifically, we have the following research objectives: 
\begin{enumerate*}[label=\roman*),font=\itshape]
   \item to quantify the presence and relationships of the social bot accounts and analyze behavioral differences with the human-controlled accounts;
   \item to infer and analyze the political party affinity of the social bots;
   \item to analyze the temporality of the social bot activity associating it with real world events; and
   \item to perform content sentiment analysis of the tweets.
\end{enumerate*}

To accomplish these objectives, the paper at hand is structured as follows: 
\begin{enumerate*}[label=\roman*),font=\itshape]
   \item Section~\ref{sec:background} will provide both the political context (Section~\ref{sec:background:politicalContext}) and a review of the literature (Section~\ref{sec:background:sota});
   \item Section~\ref{sec:methodology} will provide the required information regarding the followed methodology and the mathematical background;
   \item Section~\ref{sec:experiments} will present the results of the analysis which are later commented in 
   \item the discussion section (\ref{sec:discussion}); finally, 
   \item Section~\ref{sec:conclusions} will conclude and examine potential research lines.
\end{enumerate*}

\section{Background}\label{sec:background}
For the development and understanding of the case study it is necessary to have a basic notion of the Spanish political context, discussed in Section~\ref{sec:background:politicalContext}. Meanwhile, we comment on how this type of problems have been tackled in Section~\ref{sec:background:sota}.

\subsection{Political Context}\label{sec:background:politicalContext}
On November \nth{10}, 2019, the Spanish general election was held, where five dominant political parties (among several others) participated, namely: United We Can (\textit{Unidas Podemos}, \textsf{UP}, left-wing to far-left), the Spanish Socialist Worker's Party (\textit{Partido Socialista Obrero Espa\~nol}, \textsf{PSOE}, centre-left), Citizens (\textit{Ciudadanos}, \textsf{CS}, centre to centre-right), the People's Party (\textit{Partido Popular}, \textsf{PP}, centre-right to right-wing) and VOX (\textsf{VOX}, right-wing to far-right). 

Throughout the time window analyzed in this work, a number of remarkable events should be highlighted:

\begin{itemize}
    \item \textit{October \nth{10}, 2019} -- Santiago Abascal (\textsf{VOX}'s political leader) participated in the live national TV show ``El Hormiguero''
    \item \textit{October 19-\nth{20}, 2019} -- Riots in Catalonia
    \item \textit{October \nth{24}, 2019} -- Exhumation of Spanish fascist dictator Francisco Franco
    \item \textit{November \nth{4}, 2019} -- Electoral debate on national TV
    \item \textit{November \nth{10}, 2019} -- General election day
\end{itemize}

\begin{figure}[h]
    \centering
    \includegraphics[width=0.9\columnwidth]{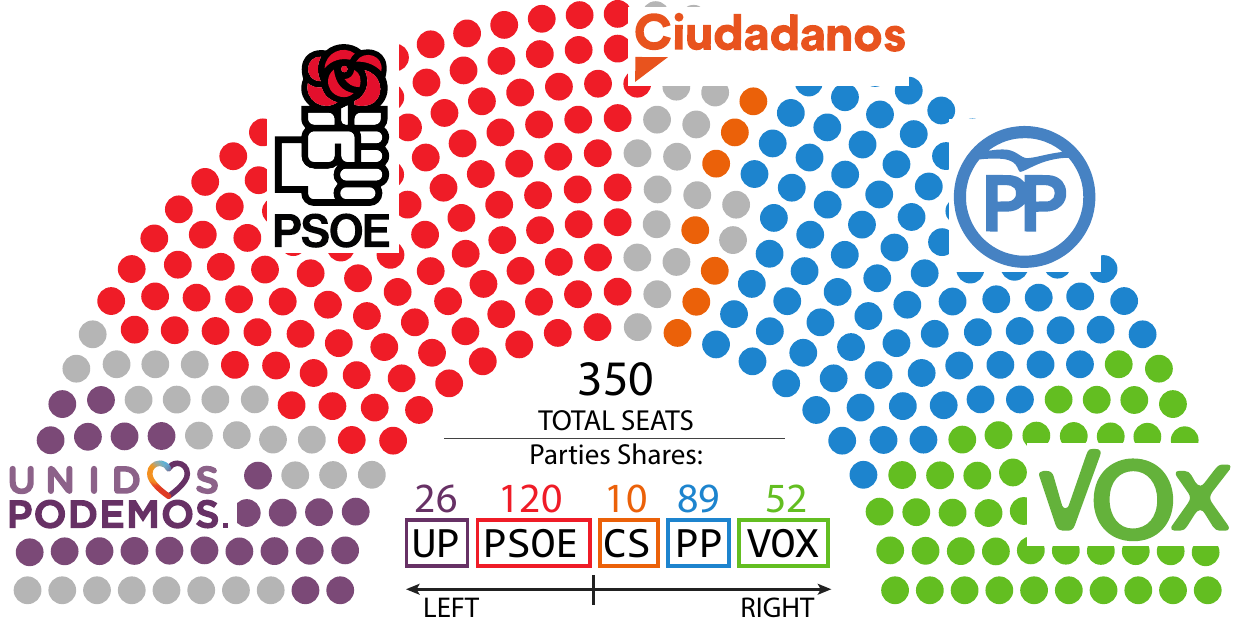}
    \caption{Parliament composition and parties' disposition after the 2019 Spanish general election, focusing on the main five political actors}.
    \label{fig:parliament}
\end{figure}

Finally, \figureautorefname~\ref{fig:parliament} reports the parliament composition as a result of the general election, note that only the five main parties are highlighted.

\subsection{Related Works}\label{sec:background:sota}
As already mentioned, social bots persist as a severe threat against modern digital democracies, attracting the attention of both academia and industry. Several researchers worldwide studied such alarming phenomenon, aiming at shedding light and proposing effective solutions.

To battle against the bots' army, authors in \cite{Varol2017} first introduced Botometer (previously known as BotOrNot), the \textit{de-facto} standard in terms of bot identification in Twitter, as detailed in Section~\ref{sec:methodology:knowledgesupervised}. By extracting a large collection of data and meta-data of Twitter users, Botometer yields quite high performance by achieving 0.85 Area Under the receiver operating characteristic Curve (AUC) when tested on publicly-available bot dataset. To increase its performance, authors added manually-labeled accounts to spot more sophisticated bots, thus reaching 0.95 AUC. Additionally, authors claimed that the bot population in Twitter ranged between 9\% and 15\%, with different retweet and mention strategies when interacting with human or other bots. Botometer has been further refined in the last years, and recently a third version has been released~\cite{Yang2019} featuring some improvements aiming at balancing the bot score.

In~\cite{Kusen2020}, the authors compared the communication patterns among Twitter users during riot events. In particular, they studied the emotional level of the messages, with a particular focus on the direct messages generated by bot accounts. Remarkably, bots convey emotions that are comparable to those conveyed by human accounts. Such a strategy is indeed used by the bots to remain unveiled and, thus, attract humans that are encouraged to interact with automatic-generated content. Similarly, in~\cite{Schuchard2019}, the relative importance and persistence of social bots were analyzed during 3 crucial events, namely: 2016 US presidential elections, the ongoing Ukrainian-Russian conflict, and the Turkish censorship implemented by the government. More specifically, the primary outcomes of the study showed that bots attempted to initiate contact with users at an extremely higher rate than human users. Through the application of social network analysis centrality measurements, authors found out that social bots, while representing less than 1\% of the total user population of the dataset, displayed an incredible level of structural network influence, with bot influencers capable of effortlessly captivating human ones.

\begin{table*}[h!]
\newcolumntype{C}[1]{>{\centering\let\newline\\\arraybackslash\hspace{0pt}}m{#1}}
%\color{blue}
\begin{tabular}{|| C{2cm} | C{2.5cm} | C{2cm} | C{1.1cm} | C{1.6cm} | C{2cm} | C{1.6cm} | C{1.8cm} ||} 
 \hline
 \textbf{Related work} & \textbf{Scenario} & \textbf{Language} & \textbf{Platform} & \textbf{Recollection period} & \textbf{Bot Detection} & \textbf{Bot presence} & \textbf{Bot activity} \\
\hline \hline
Varol et al.~\cite{Varol2017} & N.A. & English & Twitter & 90 days & Botometer threshold = 0.5 & 1.2-2.1M (9\%-15\%) & 378M-630M (9\%-15\%) \\
\hline
Kusen et al.~\cite{Kusen2020} & 2017-18 Riot Events & English, German & Twitter & 44 days & Botometer threshold = 0.6 & 9,548 (0.56\%) & N.A. \\
\hline
Schuchard et al.~\cite{Schuchard2019} & 2016 Global Events & English, Turkish, Ukrainian, Russian & Twitter & 84 days & DeBot & 14,386 (0.29\%) & 3.46M (12\%)\\
\hline
Ferrara~\cite{Ferrara2019} & 2016-17-18 Political Events & English, French & Twitter & 92 days & Botometer threshold = 0.5, simplified Botometer & 1M (18\%) & 6.525M (16\%) \\
\hline
Cresci et al~\cite{Cresci2019better} & 2012 IT Political Events, 2011 Spambots & English, Italian & Twitter & N.A. & Adversarial Approach & N.A. & N.A. \\
\hline
Pozzana, Ferrara~\cite{Pozzana2020} & 2017 FR Presidential, 2011 Spambots & English, French, Italian & Twitter & 14 days & Botometer threshold = 0.53, Hand Labeled & 2M (18\%), 5,000 (58.8\%) & 3.4M (21.25\%), 11.8M (28.8\%) \\
\hline
Luceri et al.~\cite{Luceri2020} & 2016 US Presidential & English, Russian & Twitter & N.A. & IRL, Hand Labeled & 1,148 (0.09\%) & 1.2M (8.8\%) \\
\hline
\textbf{Our contribution} & 2019 ES General & Spanish & Twitter & 41 days & Botometer threshold = 0.69, Hand Labeled & 40,098 (5\%) & 178,502 (3\%) \\
\hline
\end{tabular}
\caption{Comparative table of the analyzed related works}
\label{table:sota}
\end{table*}

Alarmingly, the social bots activity emphasizes its effectiveness during the political election campaigns, trying to deceive worldwide citizens toward forged viral trends. In~\cite{Ferrara2019}, an overview of the bots' activities during crucial political elections (\ie 2016 US presidential, 2017 French presidential, and 2018 US midterm) is presented, highlighting on some significant peculiarities. More specifically, the author uncovered massive participation of such accounts, which showed human-like behavior while performing social interactions, and, dangerously, the capability of adapting themselves to remain obscure to the Twitter detection modules. As suggested in other relevant articles, the author claimed that a black-market for political bots might exist. Recently, sophisticated detection methodologies have been proposed to win this arm race against the bot collusion as an ultimate goal. In particular, in~\cite{Cresci2019better}, authors proposed an adversarial approach to detect evolving bots. To this extent, authors synthetically modify existing bots leveraging genetic algorithm, and demonstrated the effectiveness of their proposal. Nonetheless, one could argue that the generality of such detection methodology is questionable due to the crafted nature of the algorithmically generated bots. Moreover, authors in~\cite{Pozzana2020} proposed a bot detection methodology based on the activities of the social account in Twitter. In particular, they used the concept of user session, referring to a time windows of social network posting. Specifically, the sessions analysis showed behavioral trends in humans that are absent in bots, mainly due to the automated nature of their social activity. Furthermore, an approach based on Inverse Reinforcement Learning (IRL) aiming at capturing bot behavior and thus identify bot accounts was presented in~\cite{Luceri2020}. By leveraging 2016 US presidential election data, authors were capable of correctly classifying the majority of social bots by relying on the flow of online activity within the social platform. However, it is realistic to claim that more perspectives need to be considered in order to defeat the bots' army, such as the social network structure and the posted content, just to cite some examples. 

Table~\ref{table:sota} gives a snapshot of the analyzed related works. In particular, it is worth noticing that the entire set of works investigating the bots' labor leveraging a specific platform (\ie Twitter), even though recent researches suggest the possibility of extending the investigation to other platforms~\cite{Lui2018}. Moreover, the presence and activity of the bots substantially vary depending on the proposed scenario and the employed detection tool. Interestingly, only authors in~\cite{Schuchard2019} employed DeBot as bot detection tool, arguing that it represents a valid alternative to Botometer.

On the subject of political and opinion mining, users are more likely to be exposed to online content that is ideologically closer to their political views (\ie the echo chamber) due to both the social media algorithms and the users' tendency to form links with like-minded people (\ie homophily)~\cite{Brummette2018,Zhuravskaya2019,Stefanov2019}. In this context, it has been proved that users do not form links with the opposite political side; however, they tend to mention them more often~\cite{Zhuravskaya2019}. Previous results show how different political ideologies can be distinguished by analyzing the sentiment score towards determined subjects~\cite{Chen2017}, even though these aspect-based studies assume a known target~\cite{Kannangara2018}.

To the best of our knowledge, none of the presented research works performed an in-depth analysis of the 2019 Spanish general election. Specifically, we believe that an effort is more than necessary to clarify further the role of those forged accounts in a crucial scenario for the European geopolitical scene. To this extend, this paper contributes to the human-bot arm race by proposing a novel framework which combines the capabilities of bot supervised and unsupervised learning techniques, together with a valuable features model. Additionally, by studying the social interaction and the sentiment of the posted content, an approach to correlate the bot accounts to five main Spanish political parties is proposed, employing also the manual annotation of several accounts, as we will see in the next Sections.

\section{Research methodology}\label{sec:methodology}

\begin{figure*}[h!]
    \centering
    \includegraphics[width=\textwidth]{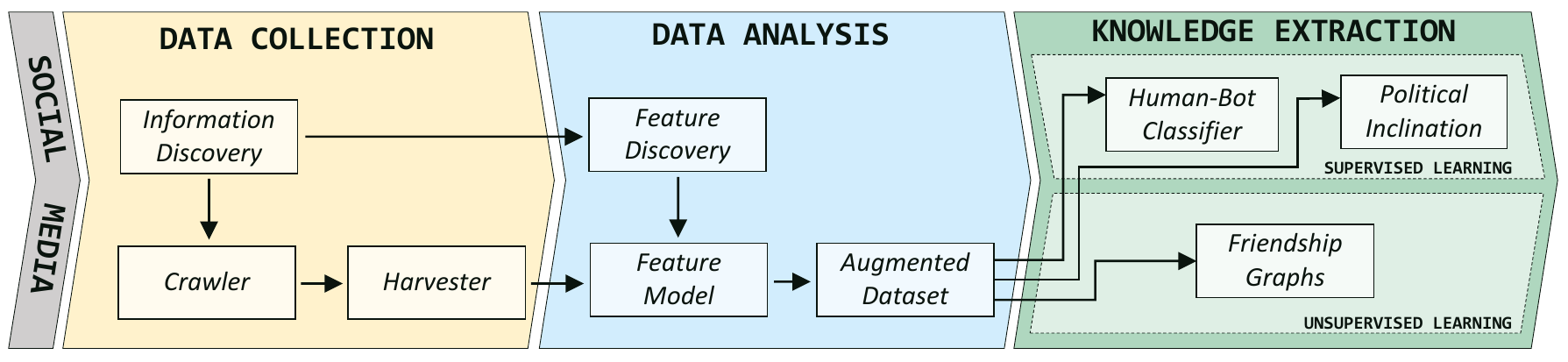}
    \caption{Research methodology adopted for social bots identification and profiling.}
    \label{fig:research-methodology}
\end{figure*}

This section illustrates the steps followed during the development of the research work. In short, this methodology is a natural evolution of the Big Data Mining for Social Bot Identification (BASTION) framework defined in~\cite{Zago2019} for social bots identification. It also evolves towards an OSINT perspective, integrating the stages defined in~\cite{Pastor-Galindo2019} for the acquisition of open-source information. \figureautorefname~\ref{fig:research-methodology} shows the followed methodology in this article, defining the modules.

As observed in the figure, the framework is composed of three main components. Specifically, the Data Collection component focuses on the acquisition of Twitter interactions using the Social Feed Manager's crawler and harvester \cite{SFM}. 
In turn, collected data goes directly to the Data Analysis component, in which the modules pick relevant features and elicit the users' identification tokens. The resulting data are stored in the Augmented Dataset for further uses. 
The core of the framework is the Knowledge Extraction component. In here, two machine learning pipelines subsequently attempt to identify and extract information from the collected data. To be more precise, firstly, a supervised ensemble algorithm classifies the Twitter accounts as either humans or bots and, secondly, a series of unsupervised techniques aim to discover correlations between the bots. 

\subsection{Data Collection}\label{sec:methodology:datacollection}

During the initial Information Discovery, we used a list of hashtags to collect tweets related to the Spanish general election. On the one hand, we initially established as hashtags in the list: the official name of the five main political parties, the associated abbreviation of the name (if existed), and their slogan of the campaign. On the other hand, along with the monitoring window, the authors paid attention to Twitter contents, searching for unpredictably trending topics related to the daily political reality. Consequently, the initial list was enriched with other hashtags regarding media scandals, conflicting situations, and important political events. Among others, authors compiled hashtags related to the visit of the political leader Abascal to the TV program ``El Hormiguero", the riots in Catalonia, the exhumation of the fascist dictator Francisco Franco, the electoral debate, or the electoral process.
The complete list, finally composed by 46 hashtags, is available in the research work's official repository~\cite{BB10N-Code}. Notice that we only collected tweets containing at least one of these hashtags.

To gather the relevant content, we deployed the Social Feed Manager (SFM) platform that continuously queried the Twitter API to accumulate the tweets (both original, retweets, replies, and quotes) containing at least one keyword within the sets mentioned above. The collection started on October \nth{4}, 2019, and concluded on November \nth{11}, 2019. During the collection window, the harvester collected around six million tweets and almost a million unique users. Despite the considerable size of the harvested data, we cannot guarantee its completeness due to limitations of the Twitter's standard search APIs. Aiming at reducing this potential gap, we recursively added all those tweets (within the observation period) that were referenced by the collected ones.The collected data is available at \cite{BB10N-Data} and explained in a published data descriptor article~\cite{BB10N-DIB}.

To better deal with the unstructured tweet data and to circumvent the relatively fixed structure of the SFM database technology, we exported the data to a non-relational MongoDB instance. We have chosen the use of MongoDB instead of complete Big Data frameworks because our volume of data is moderate, and so these solutions would generate an unnecessary increase of the complexity of our solution. At the end of the Data Collection phase, the framework had two complementary collections of data, namely the tweets ($\mathbb{T}$) and the users ($\mathbb{U}$).
To be precise:
\begin{itemize}
    \item The tweets' collection $\mathbb{T}$ stored a JSON document for each tweet containing all the objects provided by the Twitter APIs, \eg the unique identifier of the tweet, its text or the author, among others. For a full list, see both Twitter API and SFM documentation~\cite{SFM}. 
    \item The users' collection $\mathbb{U}$ stored the set of unique users collected while crawling the tweets' collection.
\end{itemize}

Throughout the following sections, we will refer to the records in these collections as named tuples and access their fields using the superscript notation. Furthermore, we will use bold $\textbf{t}\in\mathbb{T}$ and $\textbf{u}\in\mathbb{U}$ to indicate any tweet \textbf{t} or user \textbf{u}, respectively. For example, $\textbf{t}^\text{uid}$ and $\textbf{u}^\text{uid}$ indicate the unique identifier of the tweet and the user, while $\textbf{t}^\text{text}$, $\textbf{t}^\text{timestamp}$ and $\textbf{t}^\text{type}$ refer to the tweet's text, timestamp and tweet type, respectively, where $\textbf{t}^\text{type} \in \Pi=\{\text{\textsf{original}}, \text{\textsf{retweet}}, \text{\textsf{reply}}, \text{\textsf{quote}}\}$.

\subsection{Data Analysis}
The Data Analysis component takes care of transforming the raw data obtained by the Twitter API, publicly available at \cite{BB10N-Data}, into a usable format to power the machine learning pipelines. This component retrieves the full data from the harvester and, after an anonymization layer, outputs the Augmented Dataset that will be used in the analysis. 

Unveiling it, this component hosts both the feature extraction and the anonymization processes. Concerning the former, human analysts are in charge of analyzing the information retrieved from the social media and of discovering, through careful literature review, which features are required by the analysis. 

Considering our Augmented Dataset, the first and foremost augmented feature used by the framework is the tweet's sentiment score. That is, a machine learning classifier predicts a score for each tweets' text, obtaining a value ranging from zero (extremely negative) to one (extremely positive), \ie $\textbf{t}^\text{sent}\in[0,1]$. To be precise, the tweet's text also passes through a method that makes the text easier to understand (\eg removes special characters and converts emoji to text). Retweets in particular will have the same sentiment score as the original tweet.

The second augmented feature used by the framework is the tweet's topic mention. The information and Feature Discovery phases picked up five different groups of keywords ($\mathbb{W}^\Gamma$) that contained those trending topics which emerged on Twitter during the five political events highlighted in Section~\ref{sec:background:politicalContext}, namely: i) AbascalEH, ii) Catalonia, iii) exhumation, iv) debate and v) election. The Feature Model determines, for each bag-of-word $\mathbb{W}^\Gamma$, whether at least one of the contained keywords matches in the tweet's text (or original text in case of retweets).

Similarly to the previous group of features, the framework extracts whether the tweets are mentioning any political party. To be precise, the information and discovery phases identified five groups of keywords ($\mathbb{W}^P$, where $P\in\mathbb{P}=\{\text{\textsf{UP}}, \text{\textsf{PSOE}}, \text{\textsf{CS}}, \text{\textsf{PP}}, \text{\textsf{VOX}}\}$) that refer to the five major political parties in Spain. However, rather than having a generic match with these bag-of-words, these mentions to the political parties are considered exclusive. Formally, each text is labeled in a one-hot-encoding-like style, \ie a tweet labeled as \textsf{PP} only features at least one mention to the People's Party and, in the text, there is not any match with any keyword included in another political parties' bag-of-word. The same applies for the retweets.

Moreover, the framework extracts all the relations between any two users. In other words, we cave the intersections between any given user's followers and followings lists and the whole list of harvested ones.
Formally, let $\mathbb{F}_{e}$ and $\mathbb{F}_{i}$ be the followers and followings lists of the user $\textbf{u}$ provided by the Twitter's API. Then, we define ${\textbf{u}^\text{fwe} = \mathbb{F}_{e}\cap\mathbb{U}}$ and ${\textbf{u}^\text{fwi} = \mathbb{F}_{i}\cap\mathbb{U}}$ as the user's followers and followings lists respectively. 

Finally, the resulting feature set is released from any reference to the original tweets; moreover, each tweet's unique identifier is replaced with a randomly generated universal unique identifier (UUID). These changes are performed automatically, and the map between the tweets' identifiers and the newly generated UUIDs has not been saved to guarantee the unique directionality of the transformation. Note that UUIDs do not replace the users' identifiers at this point. 

After the Data Collection and Analysis stages, it was also needed to process the data to acquire useful information about all the accounts identified, being able to classify them as bots or humans, to identify the political inclination of the detected bots, and to identify relationships between bots. 

\subsection{Knowledge Extraction - Supervised Learning}\label{sec:methodology:knowledgesupervised}

\subsubsection{Human-Bot Classifier}\label{sec:methodology:knowledgesupervised:botclassifier}
To detect the presence of social bots within the dataset, we used Botometer, the \textit{de-facto} standard in terms of bot identification in Twitter~\cite{Yang2019}. This framework, developed at Indiana University, consists of a machine learning platform that extracts and analyzes more than 1200 features spanning from content-related information, user profile data, and sentiment analysis to produce a score suggesting the likelihood that the account is indeed a social bot. Botometer is publicly available through RapidAPI.

Among the several scores provided by the tool, theoretical analysis and experimental results showed that the most suitable one is the Universal Score $\textbf{u}^\text{s}\in[0,1]$, as described in~\cite{Yang2019}. It rates any account according to the likelihood of being a bot, \ie scores close to 0 indicate that the users are extremely likely to be real-user accounts, while scores nearby 1 show that the accounts are behaving like social bots. 

\begin{figure*}
    \centering
    \includegraphics[width=\textwidth]{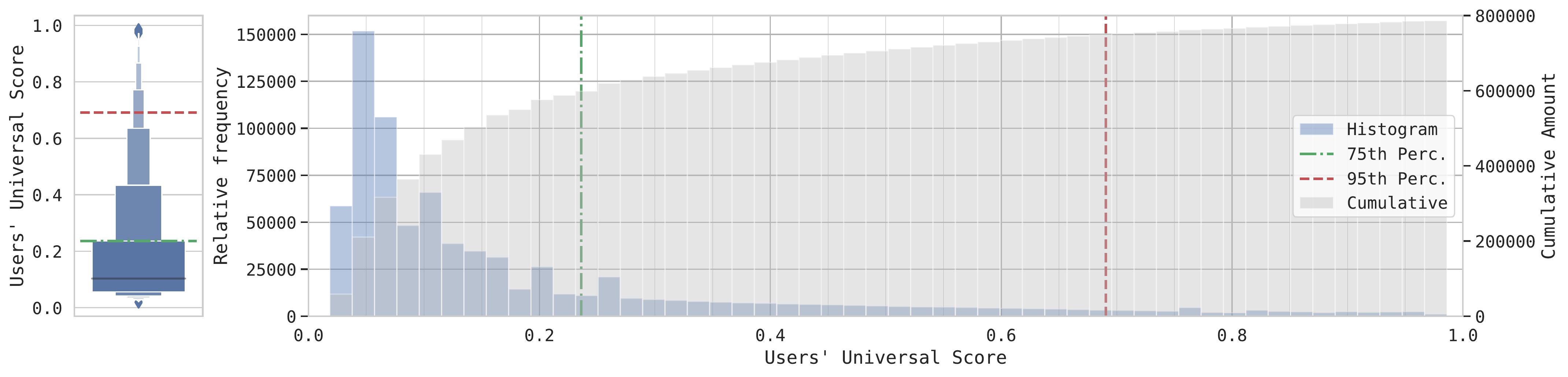}
    \caption{Users' bot score histogram, with cumulative distribution and percentiles.}
    \label{fig:percentiles}
\end{figure*}

To be as conservative as possible, the Human-Bot Classifier of our proposed framework (see \figureautorefname~\ref{fig:percentiles}) only considers the two ends of the universal score's scale for classification. That is to say, once we obtained the universal score distribution for the Augmented Dataset, a statistical approach was used to label the accounts. 
Formally, we identified with the $75^{th}$ percentile ($p_{75}\in[0,1]$) and $95^{th}$ percentile ($p_{95}\in[0,1]$) the boundaries for the human-like and the bot-like classes (establishing a percentile of the $95^{th}$ is common in the social sciences for statistical significance or detection of outliers~\cite{finlay1986statistical}). That is to say, we considered as \texttt{human} ($\textbf{h}\in\mathbb{H}\subseteq\mathbb{U}$) those users $\textbf{u}$ with $\textbf{u}^\text{s}<p_{75}$ and \texttt{bots} ($\textbf{b}\in\mathbb{B}\subseteq\mathbb{U}$) those users $\textbf{u}$ with $\textbf{u}^\text{s}>p_{95}$, labeling all the users in between as \texttt{unclear}. In our sample, with $p_{75}=0.236$ and $p_{95}=0.691$, we labeled approximately 593,000 accounts as humans, 145,000 as unclear and 40,000 accounts as social bots.

Finally, the module takes care of saving this information in the dataset and, at last, anonymizes the users' identifier, converting them into randomly generated UUIDs.

\subsubsection{Political Inclination classifier}\label{sec:methodology:knowledgesupervised:politicalclassifier}

In order to be able to solve the attribution problem, as described in~\cite{Zago2019}, the framework includes a political affiliation classifier that aims to sort the social bots according to their political behavior. As reported in~\cite{Caetano2018,Halberstam2016,Chen2017}, natural language processing (NLP) has been proved effective in analyzing the political characteristics of Twitter users. Thus, the Political Inclination classifier makes use of the average sentiment score for each group of political party's keywords ($\mathbb{W}^P$). The required overhead requested by any publicly available opinion mining framework has been deemed not effective nor necessary. To be more precise, since the data sources narrow subset of the whole population (as reported in the previous sections), classical solutions based on a plethora of dimensions do not work due to the lack of suitable data. In other words, the social bots' profiles have been analyzed only in regards to the specific political keywords identified, thus they do not present any content suitable for a more generic topic-based analysis.

Formally, the sentiment score $s_{\textbf{u},\tau,P}\in[0,1]$ for a given user~$\textbf{u}$, a tweet type $\tau$ $\in\Pi$ and a political party $P\in\mathbb{P}$ is computed as follows:

\begin{equation}
    \begin{split}        s_{\textbf{u},\tau,P}=\texttt{mean} (\textbf{t}^\text{sent} &  \in [0,1]  \ | \   \textbf{t}^\text{uid} = \textbf{u}^\text{uid} \\
       \wedge\ & \textbf{t}^\text{type} = \tau \\
       \wedge\ & \exists\ w\in \mathbb{W}^P ~.~ w\subseteq \textbf{t}^{\text{text}} \\
       \wedge\ & \forall\ P' \neq P ~.~ \forall\ w' \subseteq \mathbb{W}^\text{P'}~.~w'\nsubseteq \textbf{t}^{\text{text}})
    \end{split}
\end{equation}

It is noteworthy that, in order to calculate $s_{\textbf{u},\tau,P}$, we only consider those tweets mentioning a single political party $P$, that is to say, a tweet citing two or more parties is excluded to avoid user's opinion misinterpretation~\cite{Kannangara2018}. Moreover, such sentiment $s_{\textbf{u},\tau,P}$ is divided according to the tweet type $\tau$ to remark the differences in the type of interaction.

Furthermore, for each one of the identified subjects $\mathbb{W}^\Gamma$, this module extracts the average sentiment score depending on both the user, the tweet type, and the mentioned political party. In the same way as the previous equation, tweets with multiple political parties' mentions are not included in the analysis. However, this condition does not apply to the subject mentions, that is to say, a tweet is considered if it includes at least, but not limited to, one keyword of that subject.
Thus, formally, given a user $\textbf{u}$, a tweet type $\tau\in\Pi$, a political party $P\in\mathbb{P}$, and a subject $\gamma\in\Gamma$, the sentiment score $s_{\textbf{u},\tau,P,\gamma}\in[0,1]$ is computed as follows:

\begin{equation}
    \begin{split}
        s_{\textbf{u},\tau,P,\gamma}=\texttt{mean} (\textbf{t}^\text{sent} & \in [0,1] \ | \  \textbf{t}^\text{uid} = \textbf{u}^\text{uid} \\
        \wedge\ &\textbf{t}^\text{type} = \tau \\
        \wedge\ & \exists\ w\in \mathbb{W}^P ~.~ w\subseteq \textbf{t}^{\text{text}} \\
        \wedge\ & \forall\ P' \neq P ~.~ \forall\ w' \subseteq \mathbb{W}^\text{P'}.~w'\nsubseteq \textbf{t}^{\text{text}} \\
        \wedge\ & \exists\ q\in \mathbb{W}^\gamma ~.~ q\subseteq \textbf{t}^{\text{text}})
    \end{split}
\end{equation}

Thus, combining the equations mentioned above, we obtain a feature vector $\textbf{x}^\textbf{u}$ for each user $\textbf{u}$ that includes both the average sentiment score towards any given political party $s_{\textbf{u},\tau,P}$ and toward any subject thematic in combination with any political party $s_{\textbf{u},\tau,P,\gamma}$. Formally:

\begin{equation}
    \begin{split}
        \textbf{x}^\textbf{u} = &\big\{s_{\textbf{u},\tau,P} ~~|\ \forall\ \tau \in \Pi~.~
        \forall\ P \in \mathbb{P}
        \big\} \\ 
        \cup~&\big\{s_{\textbf{u},\tau,P,\gamma} |\ \forall\ \tau \in \Pi~.~
        \forall\ P \in \mathbb{P}~.~\forall\ \gamma \in \Gamma
        \big\} 
    \end{split}
\end{equation}

Regarding the training and testing dataset used for building the machine learning classifier, a total of 1,000 among the leading and most important verified politicians have been manually labeled with their political party~\cite{Brummette2018}. Those accounts were either verified by Twitter or explicitly mentioned the affiliation with a political party in their name or description. Such handcrafted subset of users, intentionally balanced with 200 politicians per political party, provided the training sample of the classifier ($\mathbb{X}$). 

As we did not have specific requirements for adopting a certain kind of classification algorithms, we selected six of the most common ones to perform the analysis, namely:

\begin{itemize}
    \item Random Forest ($f_1=\text{rf}(\mathbb{X})$) -- 10 trees, with  minimum leaf size of 5
    \item Multilayer perceptron - NN ($f_2=\text{nn}(\mathbb{X})$) -- with a single hidden layer, 100 nodes, ReLu activation function, Adam solver with $\alpha=0.001$, and 200 replicable training interactions
    \item Support Vector Machine - SVM ($f_3=\text{svm}(\mathbb{X})$) -- $C=1.0$, $\epsilon=0.1$ with RBF kernel, and 100 interaction limits with $0.001$ numerical tollerance
    \item Naive Bayes ($f_4=\text{nb}(\mathbb{X})$)
    \item k-Nearest Neighbor - kNN ($f_5=\text{knn}(\mathbb{X})$) -- with 5 neighbors, euclidean metric, and uniform weights
    \item AdaBoost (with internal tree) ($f_6=\text{ab}(\mathbb{X})$) -- with 50 estimators, SAMME.R classification algorithm and linear regression loss function
\end{itemize}

The evaluation of the algorithms has been carried out with a 10-fold cross validation whose results are available in \tableautorefname~\ref{table:classifier}.

\begin{table}[!ht]
    \centering
    \begin{tabular}{|l|c|c|c|c|c|}
        \hline
        \textbf{Model}   & \textbf{Accuracy}  & \textbf{Precision} & \textbf{Recall}  & \textbf{F1}    & \textbf{AUC}   \\\hline\hline
        RF      & \textbf{0.962} & \textbf{0.963} & \textbf{0.962} & \textbf{0.962} & 0.995 \\\hline
        NN      & 0.955 & 0.955 & 0.955 & 0.955 & 0.993 \\\hline
        SVM     & 0.932 & 0.934 & 0.932 & 0.932 & 0.991 \\\hline
        NB      & \textbf{0.962} & \textbf{0.963} & \textbf{0.962} & \textbf{0.962} & \textbf{0.998} \\\hline
        kNN     & 0.940 & 0.943 & 0.940 & 0.940 & 0.977 \\\hline
        AB      & 0.954 & 0.955 & 0.954 & 0.954 & 0.994 \\\hline
        
    \end{tabular}
    \caption{Evaluation of the trained classifiers with a manually labeled sample.}
    \label{table:classifier}
\end{table}

We indicate with $f(\textbf{x})=[\textbf{\^{y}},p(\textbf{\^{y}})]$ the result of applying a classifier $f$ to any given feature vector $\textbf{x}$, resulting in a set of predicted classes $\textbf{\^{y}}$ with their relative probability $p(\textbf{\^{y}})$. It holds that the probability of any given class $\textbf{y'}$ is given by ${f(\textbf{x})[\textbf{y'}]=p(\textbf{y'})}$. If the evaluated user has a manually verified label, this will be indicated with $\textbf{y}$.

\begin{figure*}[!htb]
    \centering
    \includegraphics[width=\textwidth]{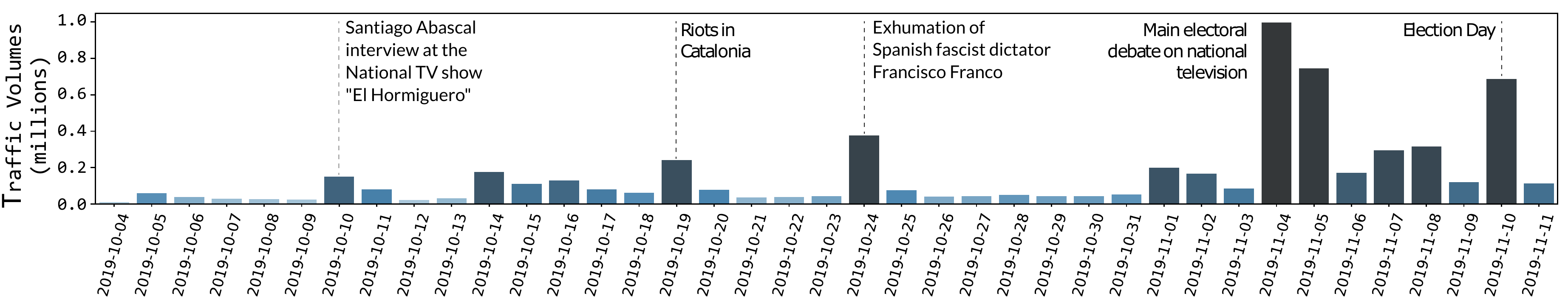}
    \caption{Tweets volumes per day.}
    \label{fig:dailyvolumes}
\end{figure*}

Given the high scores achieved by all the algorithms, we decided to combine their predictions (weighted accordingly) to label the social bot accounts. 
Formally, for a party $P\in\mathbb{P}$ and a user $\textbf{u}$ with feature vector $\textbf{x}^\textbf{u}$, the probability of being associated to that political party, $\textbf{u}\mapsto P$, is expressed as:

\begin{equation}
    p(\textbf{u} \mapsto P) = \underset{\forall\ f \in \mathbb{F}}{\text{mean}}\Big(f(\textbf{x}^{\textbf{u}})[P]\Big)
\end{equation}

In other words, the probability of the predicted label is assigned by averaging the probability assigned by each algorithm. We map an account with a political party if and only if $p(\textbf{u} \mapsto P) > \delta$, where $\delta\in[0,1]$ represents a threshold defined as:
\begin{equation}
    \delta=1-\frac{1}{|\mathbb{P}|}=0.8
\end{equation}

However, if the predicted probability for a single party is lower than $\delta$, then we consider the cumulative sum of the two parties with the highest probability. 
For example, consider a bot $\textbf{b}\in\mathbb{B}$ that has $p(\textbf{b} \mapsto \textsf{PSOE})=0.5$ and $p(\textbf{b} \mapsto \textsf{UP})=0.45$. Individually, neither \textsf{PSOE} nor \textsf{UP} reach a good enough confidence to justify the classification. However, when considered together:
\begin{equation*}
    p(\textbf{b} \mapsto \{\textsf{PSOE}, \textsf{UP}\}) = p(\textbf{b} \mapsto \textsf{PSOE}) + p(\textbf{b} \mapsto \textsf{UP}) = 0.95
\end{equation*}
Then, their cumulative sum is greater than the threshold, hence the predicted class ``\textsf{PSOE-UP}'' can be accepted.

Finally, we reject the predicted class for any bot for which it is not possible to predict a class with confidence score higher than $\delta$.

\subsection{Knowledge Extraction - Unsupervised Learning}\label{sec:methodology:knowledgeunsupervised}

The mission of the unsupervised machine learning component is to help identify, in a visual way, groups of social bots by analyzing their properties as a social group. Indeed, one of the most critical challenges~\cite{Zago2019} resides in the visualization of these large graphs. This research makes use of the Gephi software to generate an undirected graph containing the friendship relations between the studied set of bots. In particular, each node of the graph represents a unique user, while the edges (or connections) between nodes are constructed using the followers ($\textbf{u}^\text{fwe}$) and followings ($\textbf{u}^\text{fwi}$) lists of these users. More specifically, two nodes will be connected if there is a follower or following relationship between them.

We apply Gephi's ForceAtlas2 algorithm to construct the graph layout \cite{ForceAtlas2.Gephi.2014}. This algorithm constantly iterates until stopped, dynamically repulsing or attracting the nodes in a network based on the current state of the graph and a number of input parameters. Additionally, we have computed the closeness centrality of each node and used it to codify its size, where those nodes with a higher centrality will present a larger size. Finally, we keep only the giant component of the network, thus removing small and isolated sub-components of the network.
\section{Experiments and Results}\label{sec:experiments}

This section introduces the experiments and the results obtained by analyzing the 2019 Spanish general election.
All these experiments were conducted in a dedicated server featuring 2 Intel Xeon E5-2630 v4 CPUs (a total of 20 cores at 2.2 GHz) and 80 GB of DDR4 memory at 2400 MHz. The whole project occupies around 115 GB of storage.

\subsection{Statistical information}

This section presents some descriptive statistics regarding the dataset size and composition.

\begin{table}[!h]
    \centering 
    \begin{tabular}{|l|r|r|r|}
        \hline
         \multicolumn{1}{|c}{\textbf{Tweet Type}}  & 
         \multicolumn{1}{|c|}{\textbf{Amount}} & 
         \multicolumn{1}{c|}{\textbf{Proportion}} &
         \multicolumn{1}{c|}{\textbf{User's AVG}} \\\hline
        \hline 
        Retweet     & 5,116,265 & 87.81\% & 7.13 \\\hline
        Original    &   593,794 & 10.19\% & 3.75 \\\hline
        Reply       &    66,032 &  1.13\% & 3.23 \\\hline
        Quote       &    50,564 &  0.87\% & 2.33 \\\hline
        \hline
        \textbf{Total} & \textbf{5,826,655} & 100\% & - \\\hline
    \end{tabular}
    \caption{Tweet types distributions of collected tweets.}
    \label{table:interactions}
\end{table}

Firstly, \tableautorefname~\ref{table:interactions} reports the overall amounts of collected tweets according to their type. That is to say, over the $5,826,655$ tweets collected, the vast majority ($87.81\%$) are retweets, followed by originals ($10.19\%$), replies ($1.13\%$), and quotes ($0.87\%$). The table also includes the average (AVG) number of interactions per user.

\begin{table}[!h]
    \centering
    \begin{tabular}{|l|r|r|r|}
        \hline
         \multicolumn{1}{|c}{\textbf{User Group}}  & 
         \multicolumn{1}{|c|}{\textbf{Amount}} & 
         \multicolumn{1}{c|}{\textbf{Proportion}} &
         \multicolumn{1}{c|}{\textbf{AVG Tweets}} \\\hline
        \hline
        Removed     &   5,322 &  0.68\% & - \\\hline
        Humans      & 592,909 & 75.70\% & 7.36 \\\hline
        Uncertain   & 144,856 & 18.50\% & - \\\hline
        Social bots &  40,098 &  5.12\% & 4.45 \\\hline
        \hline
        \textbf{Total} & \textbf{783,185} & \textbf{100\%} & -\\\hline
        \end{tabular}
    \caption{Collected users' groups distributions.}
    \label{table:users}
\end{table}

Secondly, from the tweets we extrapolated the set of users ($783,185$ unique accounts to be precise) as indicated in \tableautorefname~\ref{table:users}; however, a number of accounts ($0.68\%$) have been excluded \textit{a-priori} either because they had been removed, they had no tweets, or they had a private profile. As firstly illustrated in \figureautorefname~\ref{fig:percentiles} and then presented in \tableautorefname~\ref{table:users}, a total of $592,909$ have been classified as \textit{Humans} (\ie $\textbf{u}^\text{s}<p_{75}$), $144,856$ as \textit{Uncertain} (\ie $p_{75}\leq \textbf{u}^\text{s}< p_{95}$), and $40,098$ as \textit{Bots} (\ie $\textbf{u}^\text{s}\geq p_{95}$). To reduce the bias in our investigations, the \textit{Uncertain} group has been discarded.

\begin{table}[!h]
    \centering
    \begin{tabular}{|l|c|r|r|r|}
        \hline
         \multicolumn{1}{|c}{\textbf{Users' Class}}  & 
         \multicolumn{1}{|l|}{\textbf{Tweet Type}} & 
         \multicolumn{1}{|c|}{\textbf{Amount}} &
         \multicolumn{1}{c|}{\textbf{Proportion}} &
         \multicolumn{1}{c|}{\textbf{User's AVG}} \\\hline
        \hline
        \multirow{4}{*}{Humans}  
          & Retweet    	& 3,756,324 & 86.29\% & 6.89 \\\cline{2-5}
          & Original   	&   497,814 & 11.44\% & 3.84 \\\cline{2-5}
          & Reply  	    &    55,116 &  1.27\% & 3.28 \\\cline{2-5}
          & Quote  	    &    43,553 &  1.00\% & 2.30 \\\hline
        \multicolumn{2}{|l|}{\textbf{Subtotal}} & \textbf{4,352,857} & \textbf{100\%} & \textbf{6.20} \\\hline
        \hline
        \multirow{4}{*}{Social bots}
          & Retweet    	&  164,927 & 92.40\% & 4.58 \\\cline{2-5}
          & Original   	&   12,444 &  6.97\% & 2.30 \\\cline{2-5}
          & Reply  	    &      659 &  0.37\% & 1.81 \\\cline{2-5}
          & Quote  	    &      472 &  0.26\% & 1.95 \\\hline 
        \multicolumn{2}{|l|}{\textbf{Subtotal}} & \textbf{178,502} & \textbf{100\%} & \textbf{4.45} \\\hline
        \end{tabular}
    \caption{Distribution of tweets per category.}
    \label{table:userstweettype}
\end{table}

Thirdly, as shown in \figureautorefname~\ref{fig:dailyvolumes}, the collected tweets are not uniformly distributed across the observation period. In the figure, we highlighted several significant political events that are associated with these traffic spikes.

Finally, the traffic volumes, their relative proportions, and the average number of interactions per user are presented in \tableautorefname~\ref{table:userstweettype} according to both the users' class and the tweet types.

\begin{figure}[H]
    \begin{subfigure}{\columnwidth}
    \centering
        \includegraphics[width=\columnwidth]{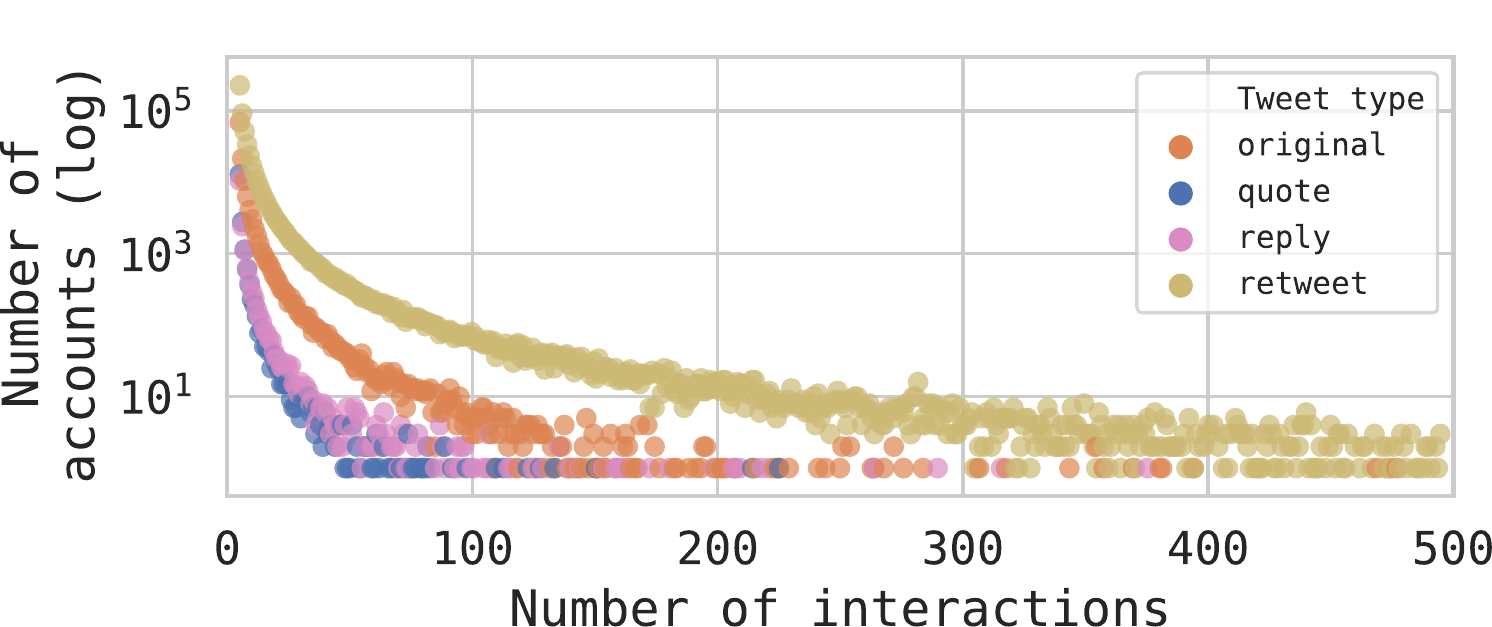}
        \caption{Distributions for those accounts classified as humans.}
        \label{fig:interactions:humans}
    \end{subfigure}\vfill
    \begin{subfigure}{\columnwidth}
        \centering
        \includegraphics[width=\columnwidth]{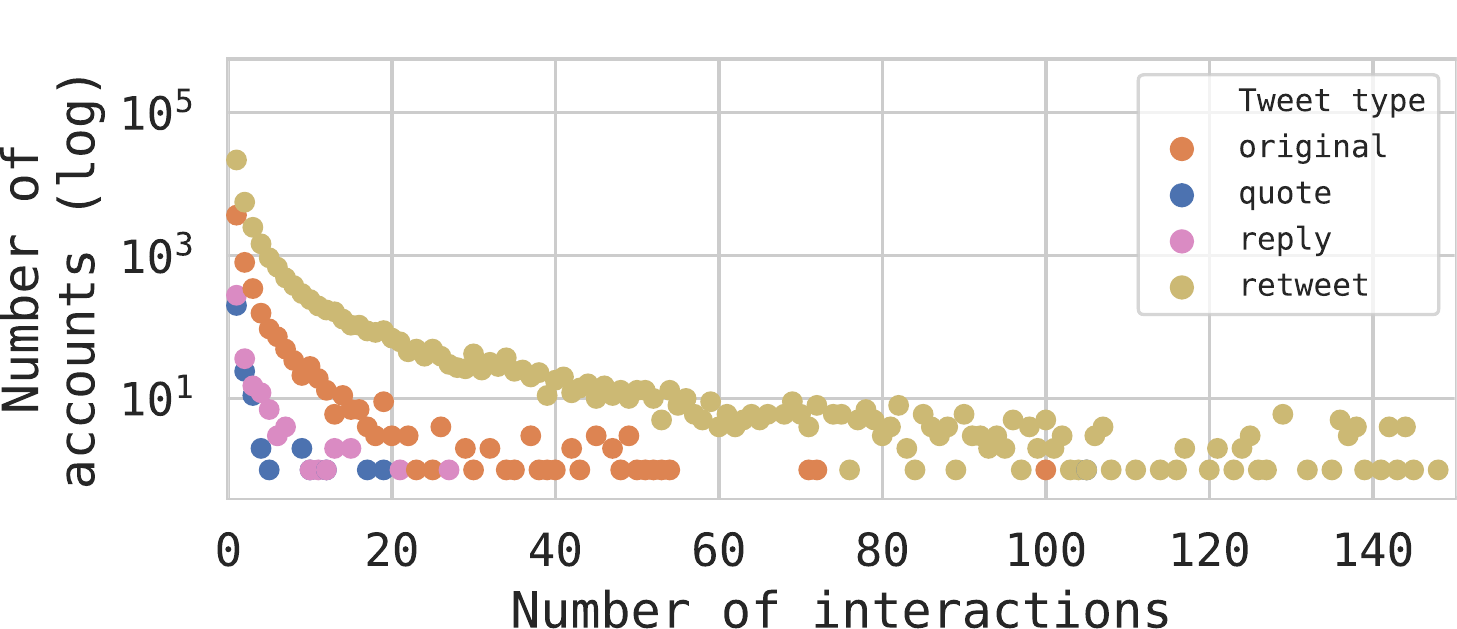}
        \caption{Distributions for those accounts classified as social bots.}
        \label{fig:interactions:bots}
    \end{subfigure}
    \caption{Number of accounts having a certain amount of interactions, grouped according to the tweet types.}
    \label{fig:interactions}
\centering
\end{figure}

In addition to these high-level metrics, we analyzed the distributions of the users according to their generated traffic volumes. In other words, \figureautorefname~\ref{fig:interactions} illustrates the number of users as a function of their number of generated interactions for those accounts classified as humans $\mathbb{H}$ (\figureautorefname~\ref{fig:interactions:humans}) and social bots $\mathbb{B}$ (\figureautorefname~\ref{fig:interactions:bots}). To improve the figure's readability, the vertical axes feature a logarithmic scale while the horizontal ones are capped both for humans and social bots. 

As shown in both \figureautorefname~\ref{fig:interactions:humans} and \figureautorefname~\ref{fig:interactions:bots}, the interactions' volumes do not present a uniform distribution; on the contrary, the vast majority of the users (both human and social bots) have published just a few tweets. Interestingly enough, all tweet types present both similar shares and distributions (although scaled), as numerically reported in \tableautorefname~\ref{table:userstweettype}.

\subsection{Behavioral differences between humans and social bots}\label{sec:4B}
One of the main questions raised by this research is to measure the effectiveness of social bots. A first attempt includes the analysis of the interactions for those tweets that have a direct and unique target, \ie retweets, replies and quotes. These tweet types indicate that a user is actively interacting with a target on different degrees. For example, a user might retweet content because of the target's idea, or comment on it (either reply or quote) due to shared interests or concerns.  

\begin{figure}[H]
    \centering
    \includegraphics[width=0.9\columnwidth]{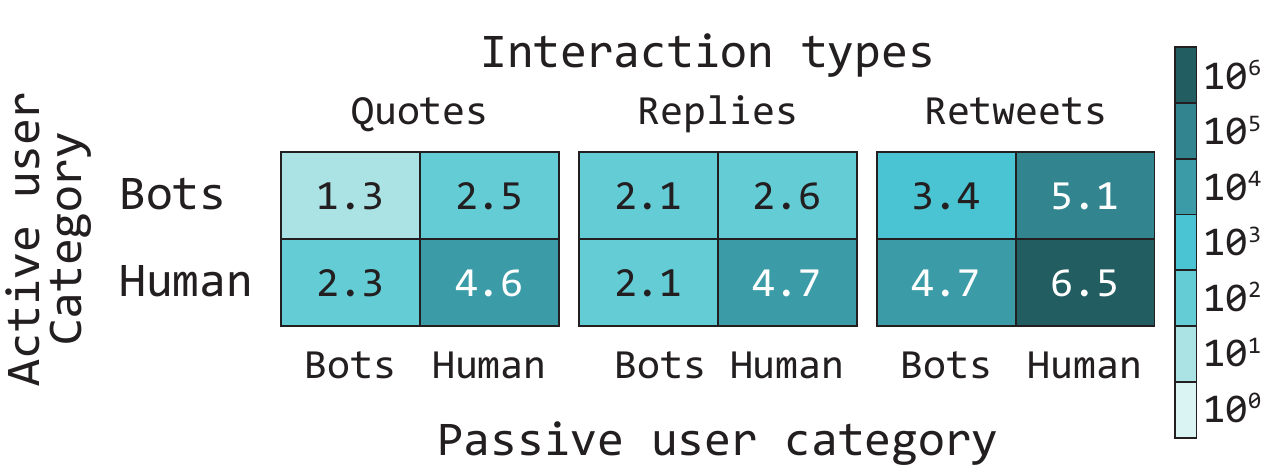}
    \caption{Interactions involving bots and users by tweet type.}
    \label{fig:volumelog}
\end{figure}

In this context, we denote as ``active user'' the person who creates the interaction whereas ``passive user'' is the owner of the retweeted, replied, or quoted tweet. Following this convention, \figureautorefname~\ref{fig:volumelog} presents the volumes of shared tweets according to both the tweet type (\ie retweets, replies or quotes) and the users' classification. Note that:
\begin{enumerate*}[label=\roman*),font=\itshape]
    \item the volumes are presented in their logarithmic form to increase the separability of the smaller categories, and, 
    \item this section only considers those tweets where both the active and the passive users are in the humans or the social bots groups; in other words, tweets starting or targeting an account in the \textit{Uncertain group} (\ie ${p_{75}\leq \textbf{u}^\text{s}< p_{95}}$) have been excluded (around $1,253,955$ tweets).
\end{enumerate*}

Nevertheless, and as expected, the single most common type of interaction is the retweet, providing alone more than $87\%$ of the interactions. Considering the whole scenario, $5\%$ of the traffic volumes involve a social bot either as an active or passive actor. However, if we exclude the human-to-human retweets, the reader might notice that this proportion skyrockets to an overwhelming $66.4\%$.

On the one hand, a different picture is depicted by looking deeper into the human's interactions other than the human-to-human retweets. Notably, only $37\%$ of these interactions are targeting social bots, of which $95\%$ are retweets. In other words, humans tend to retweet the content shared by the social bots instead of quoting or replying to them.
On the other hand, only $2\%$ of social bots activities are targeting other social bots. In particular, social bots tend to retweet human contents in an attempt to make it viral~\cite{Zago2019}: according to our data, and if we exclude human-to-human retweets, almost half of the generated traffic ($45\%$) is provided by social bots retweeting humans' contents.

\subsection{Political party affinity}\label{sec:4C}

This section focuses on the analysis of the contents shared by the social bots in an attempt to solve the attribution problem identified in~\cite{Zago2019}.
To do so, only those social bots $\textbf{b}$ that have at least one tweet specifically targeting a political party $P$ are considered. In this sense, of the nearly forty thousand social bots identified in \tableautorefname~\ref{table:users}, only $20,364$ qualifies ($50.79\%$), for a total of $104,989$ tweets ($58.82\%$).

\begin{figure}[ht]
    \centering
    \includegraphics[width=\columnwidth]{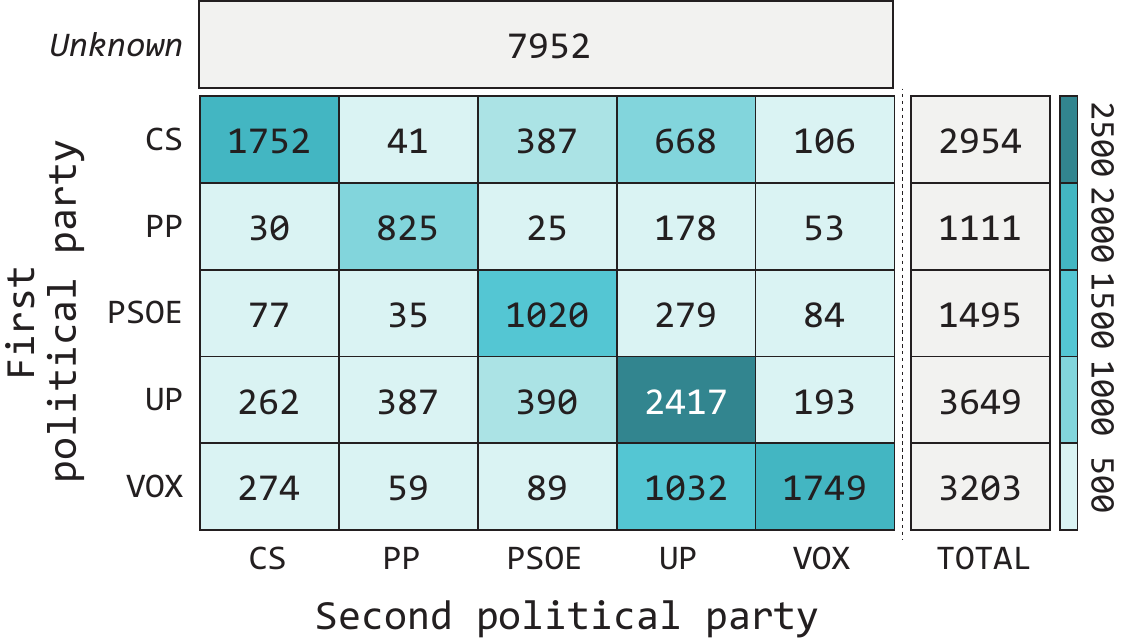}
    \caption{Number of social bots according to their predicted political party.}
    \label{fig:partiesheatmap}
\centering
\end{figure}

\figureautorefname~\ref{fig:partiesheatmap} reports the results of the ensemble classifier described in Section~\ref{sec:4C}. In the figure, the vertical axis represents the first political party identified by the classifier, while the horizontal presents the second one identified. It follows that the diagonal cells represent those social bots that have been classified as mapped to a single political party. In the figure, those social bots that have not been classified with sufficient enough confidence are labeled as ``Unknown''.

\begin{figure}[h]
    \centering
    \includegraphics[width=\columnwidth]{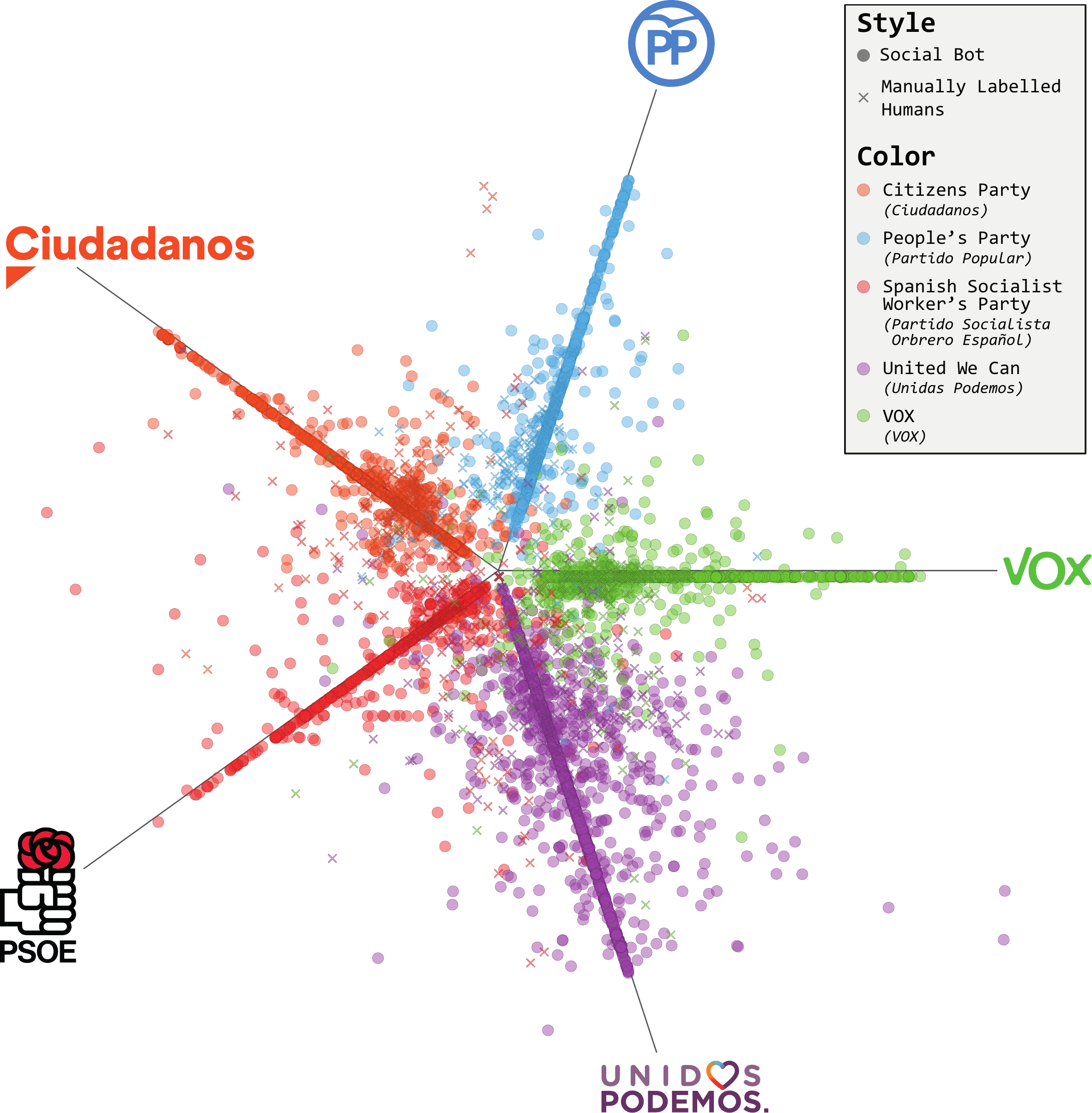}
    \caption{Bi-dimensional projection of social bots' single political affinities and manually labeled users.}
    \label{fig:projection}
\end{figure}

Classification results of those social bots aligned with only a single party are graphically represented in \figureautorefname~\ref{fig:projection} in conjunction with manually labeled users. To be precise, the figure is the result of a manually-driven dimensionality reduction to five dimensions projected into a plane. Each dimension constitutes the average user's sentiment towards a political party.

From now on, and for the sake of simplicity, we will focus the rest of the analysis on the social bots associated with only one political party.

\subsection{On the subject of social bots, followers, and followings}\label{sec:gephi}
The groups of social bots identified so far may have several common traits and could implant a coordinated behavior. Although the study of the existence of potential botnets remains as future work, it is worthy to show certain relationships between the social bots. As social media like Twitter features direct connections between the users in terms of followers and followings, it is specially interesting to place the social bots in a social graph.

\begin{figure}[h]
    \centering
    \includegraphics[width=\columnwidth]{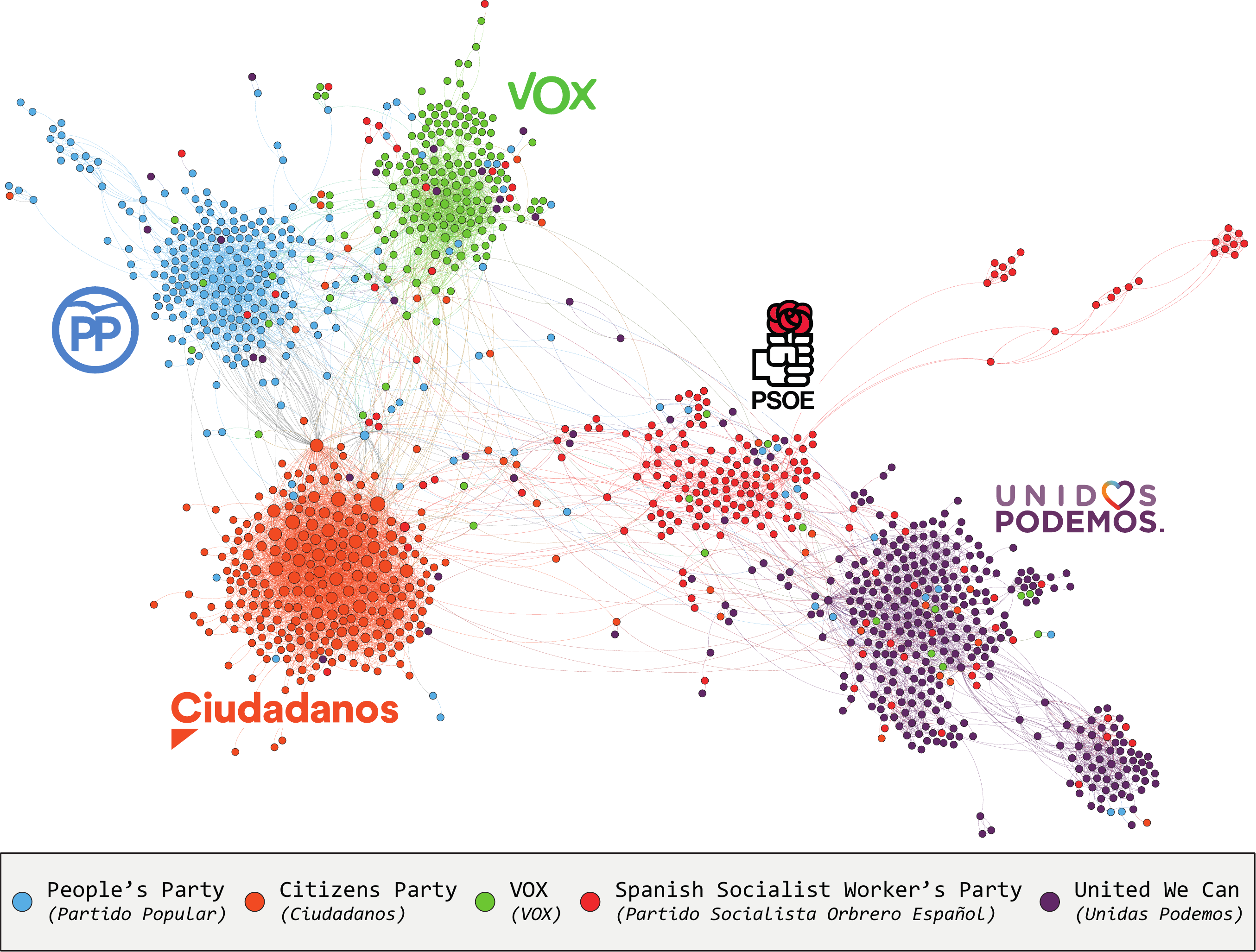}
    \caption{Friendship relationship among bots.}
    \label{fig:botnets-influence}
\end{figure}

\figureautorefname~\ref{fig:botnets-influence} presents the undirected social graph formed by the classified social bots. Each node is a social bot whereas each edge represents an undirected link between two social bots due to a following or a follower relation. 

\begin{figure*}[t]
\includegraphics[width=\textwidth]{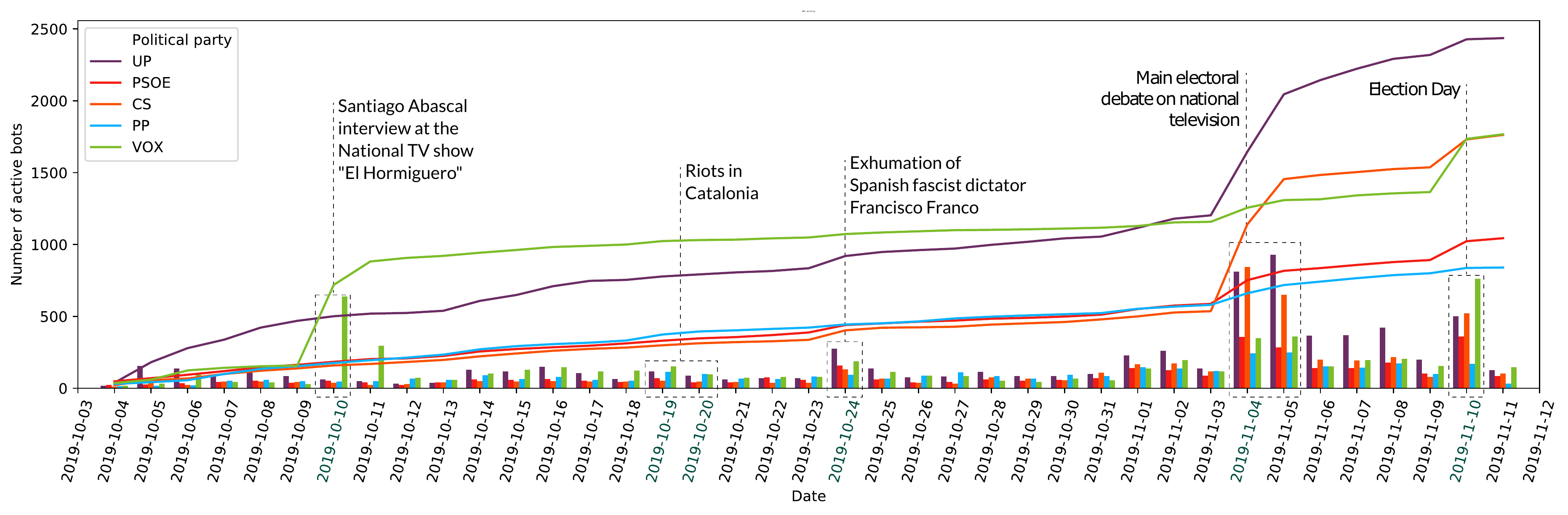}
\caption{Active social bots with one-party affinity (cumulative and per day basis).}
\label{fig:onepartyappearance}
\end{figure*}

The algorithm used to build the graph is based on network metrics such as the average degree and the centrality values of the social relationships. It places very close those nodes that are very connected to each other, and moves away those others with which fewer relationships are shared. As a result, some differentiated clusters arise, where bigger nodes indicate a higher centrality measure of the account, \ie social hubs.

Surprisingly, the resulting clusters are mostly composed by social bots that share political affinity. In other words, the groups of bots identified by the political classifier in Section~\ref{sec:4C} are indeed so socially interconnected. In combination with the quantitative measures reported in the previous sections, \figureautorefname~\ref{fig:botnets-influence} further suggests a coordinated effort to deploy, maintain, and employ social bots. 

\subsection{The temporal appearance of social bots}
\label{sec:presence}

The presence of social bots on Twitter on the eve of the 2019 Spanish general election was not random. According to our data, a direct correlation manifests between their appearance, the spikes in the traffic volumes, and the major political events.

\begin{figure*}
    \begin{subfigure}{\textwidth}
        \includegraphics[width=\textwidth]{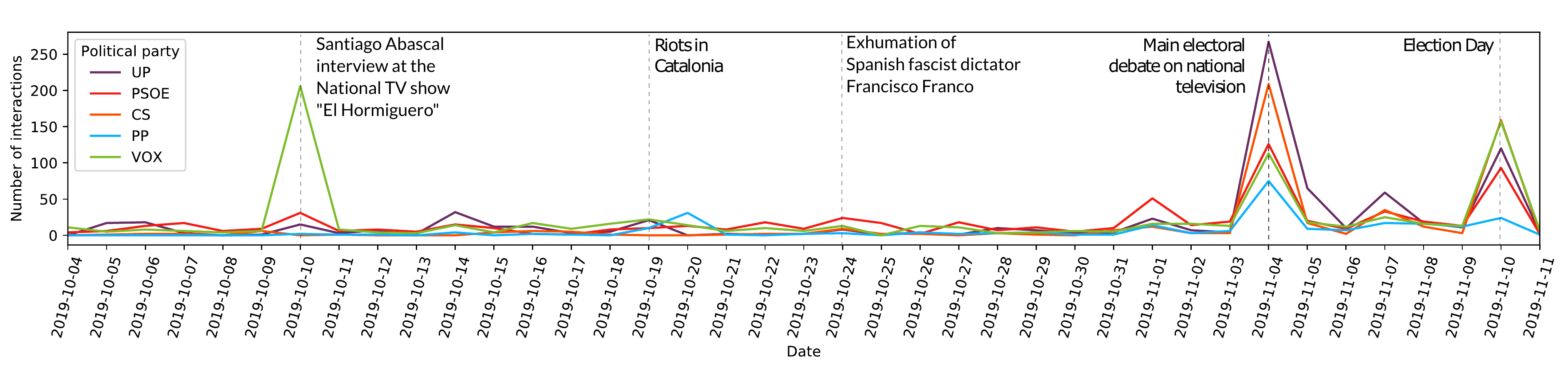}
        \caption{Originals, quotes and replies.}
        \label{fig:traffic-timeline-1}
    \end{subfigure}\vfill
    \begin{subfigure}{\textwidth}
        \includegraphics[width=\textwidth]{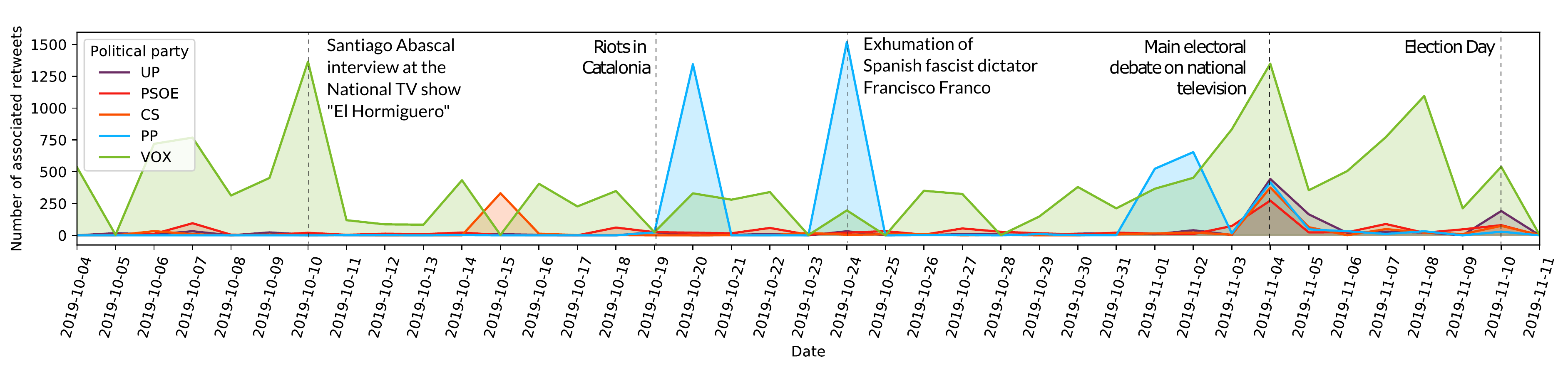}
        \caption{Associated retweets.}
        \label{fig:traffic-timeline-2}
    \end{subfigure}\vfill
    \begin{subfigure}{\textwidth}
        \includegraphics[width=\textwidth]{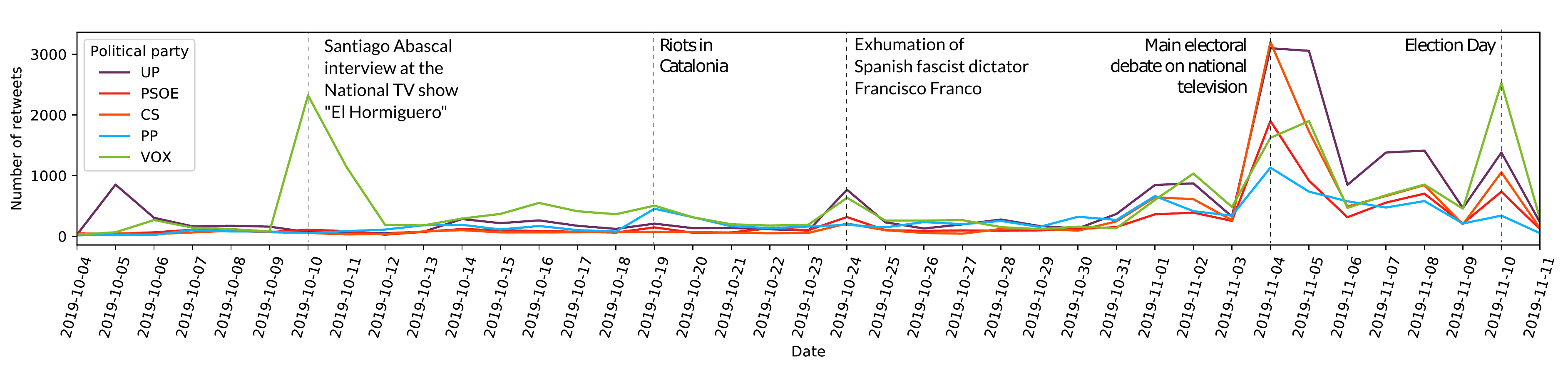}
        \caption{Total retweets.}
        \label{fig:traffic-timeline-3}
    \end{subfigure}
    \caption{Activity of social bots with one-party affinity.}
    \label{fig:traffic}
\centering
\end{figure*}

To begin with \figureautorefname~\ref{fig:onepartyappearance}, it illustrates the daily appearance of new social bots. On the horizontal axis we have the dates in the observation period, while, on the vertical one, we observe the number of detected social bots. There are two complementary groups of time series in \figureautorefname~\ref{fig:onepartyappearance}. The time series represented by the lines reports the cumulative number of unique social bots, while the histogram ones illustrate the daily activity per political party. 

From the figure, it is possible to notice that the increase in the numbers of the social armies was constant throughout the whole electoral season. Notably, more than 500 bots (exposing an affinity with \textsf{VOX}) emerged on the same day as \textsf{VOX}'s leader participated in the national TV show ``El Hormiguero''. Concerning the dates when the riots in Catalonia occurred, our data do not include any significant increase in the number of social bots. However, we did detect a sizable spike during the controversial act of Franco's exhumation. 

Notably, during the ten days before the election day, more than five thousand new social bots popped up, mainly during the night of the electoral debate and the election day. During this specific time window, two different patterns appear. On one side, the social bots associated with the \textsf{PP}, \textsf{PSOE}, and \textsf{VOX} parties presented essentially the same growing trend, while those bound to both \textsf{UP} and \textsf{Cs} depict a remarkable increment. One might infer that those fake accounts were precisely hired to support the final events of the campaign. 

Although correlation does not imply causality, it is difficult not to suspect that the social bots were driven by actors who followed, monitored, and participated in the Spanish political life. The following sections will attempt to corroborate this hypothesis by looking at the behavior of these social armies.

\subsection{Timeline of the social bots' interactions}
\label{sec:activity}

The daily increase in the number of social armies in principle does not imply their effectiveness. While, on the one hand, Section~\ref{sec:4B} presented the humans-bots interactions, on the other hand, \figureautorefname~\ref{fig:traffic} shows their temporal properties.

To be more precise, \figureautorefname~\ref{fig:traffic-timeline-1} reports, according to the political party, the number of generated original tweets, quotes and replies created by social bots. As mentioned in the previous sections, all the anomalies in the traffic volumes' patterns are correlated to major political events. As for the retweets, we devote two complementary figures. \figureautorefname~\ref{fig:traffic-timeline-2} quantifies the total retweets caused by the social bots contents, whereas \figureautorefname~\ref{fig:traffic-timeline-3} reports the volumes of the actual retweets shared directly by the social bots.

In other words, \figureautorefname~\ref{fig:traffic-timeline-2} presents the total volume of tweets that retweeted social bots original tweets, quotes and replies. Note that the overall numbers are in the orders of tens of thousands of retweets. By looking at \figureautorefname~\ref{fig:traffic-timeline-2}, it appears that the social bots allegedly connected to \textsf{VOX} promote contents that are attractive enough to be shared and propagated by other users. Oddly enough, the tweets originated from social bots affiliated with the \textsf{PP} obtained several thousand retweets on three separate occasions, namely during the riots in Catalonia, during the exhumation of the Spanish fascist dictator Francisco Franco, and in the days before the national debate. Ostensibly, these social bots shared contents aligned with the political agenda of \textsf{PP}. 

Finally, in \figureautorefname~\ref{fig:traffic-timeline-3}, it appears that, apart from a few precise exceptions, the amount of retweets during the observation period is consistently below a thousand tweets per group. Besides this steady behavior, further research is required to analyze these retweeting strategies and their interactions with other parties. 

\subsection{Tweets' contents analysis}
The social bots aspects discussed so far are mostly measures of the account's properties. However, the correlation between their actions and the subjects and ideas promoted is also of great importance.

\begin{figure*}
    \begin{subfigure}{\textwidth}
        \includegraphics[width=\textwidth]{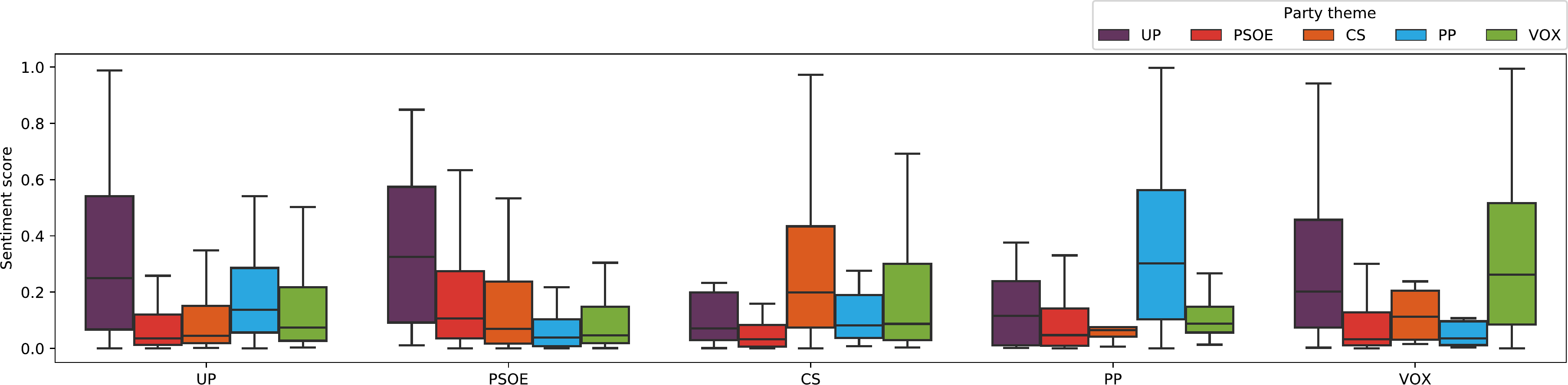}
        \caption{Manually verified human accounts.}
        \label{fig:sentimentboxplots:humans}
    \end{subfigure}\vfill
    \begin{subfigure}{\textwidth}
        \includegraphics[width=\textwidth]{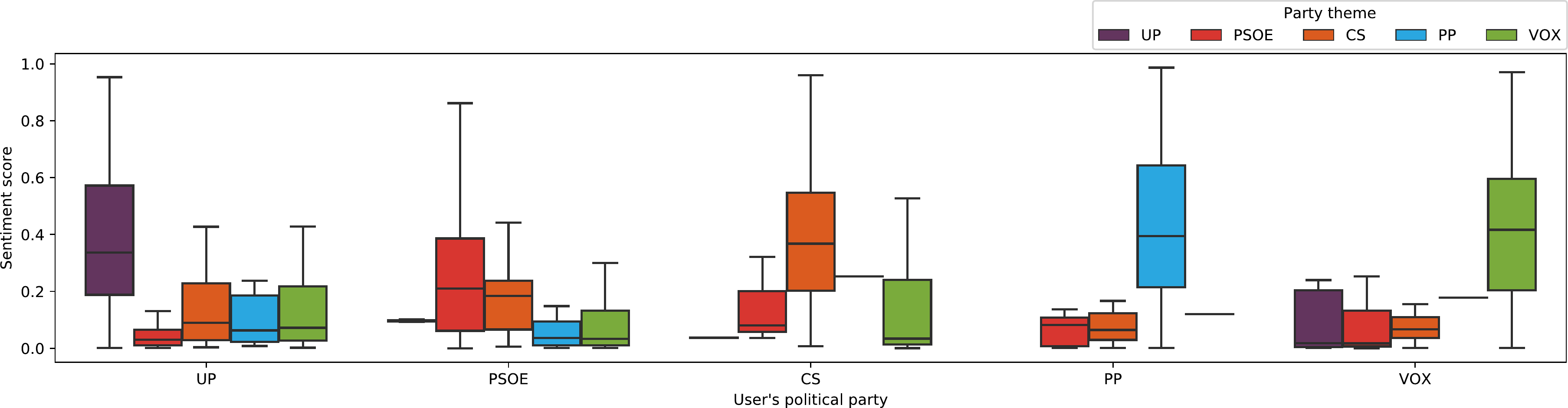}
        \caption{Social bots.}
        \label{fig:sentimentboxplots:bots}
    \end{subfigure}
    \caption{Sentiment of labeled users and social bots against party themes.}
    \label{fig:sentimentboxplots}
\centering
\end{figure*}

In this regard, \figureautorefname~\ref{fig:sentimentboxplots} reports the sentiment analysis of each group of social bots in comparison with the manually labeled sample of verified human users. The boxplots in the figure show the distribution of the sentiment score of the tweets toward each political party, \ie their attitude concerning the alliances. Both humans (in \figureautorefname~\ref{fig:sentimentboxplots:humans}) and social bots (in \figureautorefname~\ref{fig:sentimentboxplots:bots}) are divided according to their assigned political party. Scores neighboring zero indicate an antagonistic opinion, while a score close to one suggests an appraisal for the subject. 

Although the overall perspective around the parties is negative (global averages are below $0.2$), each group presents a reasonable change regarding their corresponding party. The political scenario surrounding the Spanish general election might justify the anomalies for some specific combinations. For example, in \figureautorefname~\ref{fig:sentimentboxplots:humans}, the manually verified humans representing the \textsf{PSOE} were substantially pushing more supportive content towards the \textsf{UP} than their own party. 

As for the social bots of \figureautorefname~\ref{fig:sentimentboxplots:bots}, the panorama looks appreciably different. As stated before, despite the negatively-oriented sentiment scores that characterize the political discussion, it seems that the ensemble classifier managed to separate the social bots accordingly to the supported political party.

\subsection{Overview of social bots' behavior findings}
On the basis of the results obtained, we can extract up-to-date insights of social bot behavioral patterns. In the following, we discuss the lessons learned about the properties of the monitored social bots in the context of our study:
\begin{enumerate}
    \item \textit{Social bots had a fixation on targeting major national events.} Our bot activity analysis suggests that social bots were aware of the mainstream events and sensible to trending topics. Analyzing the data of these specific situations could have an added value in detecting and monitoring social bots.
    \item \textit{Social bots tended to retweet human content.} In terms of traffic generated, social bots specifically viralized legitimate tweets rather than supporting each other or replying to other users. Even though we cannot give a founded explanation with the obtained results, we understand that the retweet is an extremely informative interaction for profiling social bots.
    \item \textit{Social bots had no affinity for a single political party, but shared more than one ideology}. From the 20,364 profiled social bots, we are only able to classify the 38\% within a unique political inclination, whereas the rest were aligned with two political parties (23\%) or more (39\%). Current social bots may be acting under more complex dynamics than supporting a single group.
    \item \textit{Social bots with the same ideology were highly socially connected}. The social graph shows that groups with the same political affinity also formed robust clusters of friendship connections on Twitter. Therefore, the position of the nodes (users) in such networks could facilitate the detection of suspicious accounts due to the proximity to clusters of bots already recognized.
    \item \textit{Social bots opted for conflict and disrepute rather than support.} The sentiment analysis demonstrates that social bot interactions were specially negative against political parties opposed to the inferred ideology. With this in mind, perhaps more consideration needs to be given to who the social bots are acting against rather who they are supporting.
\end{enumerate}
These key findings may be potentially valuable in current real-world applications, specifically, to enhance today's bot detection mechanisms. The translation of these facts into state-of-the-art methodologies and low-level machine learning features may enable the construction of more precise systems which are able to adapt and improve themselves following the latest trends of social bot behaviors.
\section{Discussion}\label{sec:discussion}

First, we would like to map the results and data that we observe with the Spanish political context at the time. This election took place due to a previous failed attempt where there was no majority in the parliament, and the \textsf{PSOE} leader did not accomplish to receive support from the rest of the parties to become Prime Minister. Hence, during the pre-electoral period, the political parties were aware of the inevitable necessity to find alliances in other parties as an absolute majority was unlikely to be reached with just the citizens' voting. \figureautorefname~\ref{fig:sentimentboxplots} showed that the overall sentiment trend in our data sample is negative, and this is aligned with the trend in many social networks and communities, where trolling and hate speech are the norm~\cite{jakubowicz2017alt_right}; however, we did report some differences in terms of sentiments as part of the interactions between some parties. For example, in \figureautorefname~\ref{fig:sentimentboxplots:humans} we see a positive sentiment score of \textsf{PSOE} towards \textsf{UP}, and not in vain, the outcome of this election was a coalition government between these two parties. Interestingly enough, the network representation of social connections presented in \figureautorefname~\ref{fig:botnets-influence} reflects the actual positioning of the political parties. To be more precise, both right-wing (\textsf{PP} and \textsf{VOX}) and left-wing (\textsf{PSOE} and \textsf{UP}) parties appear to be closely connected, using the central party (\textsf{Cs}) as a hub. Indeed, the Citizens party (\textsf{CS}) reports the highest average degree and centrality values of the network. 

Regarding the levels of activity of these accounts in Twitter, perhaps the most noteworthy finding is the very clear association between real-world events with peaks of activity of the different parties; this association between bursts of activity and media events have been further explored by previous work~\cite{lin2014rising}. Some of these events include the visit of Santiago Abascal to the national TV show ``El Hormiguero'' associated with a very high peak of activity of \textsf{VOX}, the riots in Catalonia which are associated with activity peaks of both \textsf{VOX} and \textsf{PP}, or the exhumation of the Spanish fascist dictator Francisco Franco, with a high number of original tweets from \textsf{PP}, and retweets in the case of \textsf{VOX} and \textsf{UP}. Then, on other events, such as the main electoral debate and the elections' day, the political bots of all parties were highly active in terms of interaction volumes. All of these represent key events where political parties can emphasize their views in order to gain votes, and thus this can explain the increased levels of activity at those times. These differences in how the groups of bots of each party have been orchestrated based on different events might be indicative of centralized coordination behind the scenes with clear political goals, perhaps in the form of botnets that should be further studied.

Furthermore, the study also has some limitations that we would like to acknowledge. The first one is that we are using the external tool Botometer to detect these bots, and thus our work can only be as trustworthy as this tool; however, Botometer is widely considered as the best option in the state of the art. Several decisions have been taken based on statistical and empirical measures, and our decisions lie more on the conservative side; therefore, we believe that the real number of bots and influence is significantly higher than the estimates that we report. Some uncertainty is common under this kind of computational studies in social networks where the ground truth is not directly observable. 

Once again, we would like to highlight that the research proposed in the paper at hand requires profound and critical future works on the subject of opinion mining and Spanish language processing. Moreover, despite having presented accurate classification results, we sought after an in-depth analysis of each social bot. To be precise, the highly filtered nature of the collected data does not permit to gather general information regarding the actual ideology showed by the accounts. Further researches are also needed to pinpoint sarcasm and address language-specific challenges.

Finally, we believe this line of work to be of high importance, and the rationale is grounded on psychological research on belief and behavior. Previous researches have shown how social media might be changing the political beliefs of citizens~\cite{swigger2013online}, and there have been researches attempting to model how this belief influence might propagate across social networks~\cite{nguyen2012influence}.

A number of theories have connected the beliefs and behavior of people, the latest one in~\cite{fishbein2011predicting} known as the ``Reasoned action approach,'' provides a framework that connects beliefs, intentions, and behavior. Therefore, these social bots, by importantly amplifying and propagating specific ideas, can affect the belief of the social media users, thus directly affecting their behavior when voting in political elections. It remains a challenge how to effectively quantify this effect on constituents, but we believe that controlled A/B experiments with interventions that can include awareness messages, surveys and focus groups, are promising directions to start estimating its implications. The majority of social networks including Twitter, currently are loose enough to allow certain automatization. However, given the high-stakes that are at play when these features are misused with ill objectives, we consider vital to keep studying the potential effects of these phenomena on our modern democracies.

\section{Conclusions and Future work}\label{sec:conclusions}
This paper has analyzed the presence and behavior of social bots on Twitter in the 2019 Spanish general election scenario. To summarize the article at hand, the analysis has been performed by following a research methodology composed of three main stages that encompass both supervised and unsupervised learning, namely: data collection, data analysis, and knowledge extraction. As a result, the proposed framework presents capabilities such as human and social bots classification, social bots' political inclinations identification, and a hint of social botnet discovering through friendship analysis. 

A pool of experiments has demonstrated not only a non-negligible number of social bots on Twitter participating in the Spanish elections but also a relevant number of daily interactions and traffic volume. Indeed, the analysis of behavioral differences between humans and social bots have detected that humans tend to retweet content shared by the social bots, while social bots tend to retweet human contents to make it viral. Our analysis also analyzed and reported quantitative measurements of the social bots' temporal appearance and relationships. Last but not least, although the sentiment analysis reported an overall negative trend\textemdash a common aspect in social media nowadays\textemdash it also suggested essential differences between the political parties.  

As future work, we plan to extend the harvested dataset with further interactions that would improve the performance of the classifier algorithms, perhaps including OSINT metadata regarding political actors. We also consider the revision of the sentiment analysis process by adding new metrics and improving the classification capabilities, which, to date, are limited for the Spanish language. Future studies should test the generalization of these findings, by applying similar methodologies to analyze other political election events in different countries and social networks. Finally, we will investigate the presence of potential botnets in the Spanish general election scenario as well as measure their influence in human decisions and the result of the election.

\section*{Acknowledgments}
This study was partially funded by a grant from the Spanish National Cybersecurity Institute (INCIBE) with code INCIBEI-2015-27353, by the Spanish Government grants FPU18/00304, FJCI-2017-34926 and RYC-2015-18210, co-funded by the European Social Fund, by a predoctoral grant from the University of Murcia and by the Irish Research Council, under the government of Ireland post-doc fellowship (grant code GOIPD/2018/466). Authors would also like to acknowledge Prof. Karl Aberer at EPFL, Prof. Albert Blarer at Armasuisse, H\'ector Cordob\'es and IMDEA Networks Institute for their support to this work.

\bibliographystyle{elsarticle-num}
\bibliography{biblio}
\end{document}